# Equivalence between the zero distributions of the Riemann zeta function and a two-dimensional Ising model with randomly distributed competing interactions

Zhidong Zhang

Shenyang National Laboratory for Materials Science, Institute of Metal Research,

Chinese Academy of Sciences, 72 Wenhua Road, Shenyang, 110016, P.R. China

## Abstract

In this work, we prove the equivalence between the zero distributions of the Riemann zeta function $\zeta(s)$ and the partition function of a two-dimensional (2D) Ising model with a mixture of ferromagnetic and randomly distributed competing interactions. At first, we review briefly the characteristics of the Riemann hypothesis and its connections to physics, in particular, to statistical physics. Second, we build a 2D Ising model, $M_{FI+SGI}^{2D}$, in which interactions between the nearest neighboring spins are ferromagnetic along one crystallographic direction while competing ferromagnetic/antiferromagnetic interactions are randomly distributed along another direction. Third, we prove that all energy eigenvalues of this 2D Ising model $M_{FI+SGI}^{2D}$ are real and randomly distributed as the Möbius function $\mu(n)$, the Dirichlet $L(s, \chi_k)$ function as well as the Riemann zeta function $\zeta(s)$. Fourth, we prove that the eigenvectors of the 2D Ising model $M_{FI+SGI}^{2D}$ are constructed by the eigenvectors of the 1D Ising model with phases related to the Riemann zeta function $\zeta(s)$, via the relation $\omega_1(\gamma_{2j})$ between the angle $\omega_1$ and the energy eigenvalues $\gamma_{2j}$, which form the Hilbert-Pólya space. Fifth, we prove that all the zeros of the partition function of the 2D Ising model $M_{FI+SGI}^{2D}$ lie on a unit circle in a complex temperature plane (i.e. Fisher zeros), which can be mapped to the zero distribution of the Dirichlet $\mathrm{L}(s, \chi_k)$ function and also the Riemann zeta function $\zeta(s)$ in the critical line. In a conclusion, we have proven the closure of the nontrivial zero distribution of the $\mathrm{L}(s, \chi_k)$ function (including the Riemann zeta function $\zeta(s)$).

The corresponding author: Z.D. Zhang,

Tel: 86-24-23971859,

Fax: 86-24-23891320,

e-mail address: zdzhang@imr.ac.cn

## 1.Introduction

The Riemann hypothesis is one of the most fundamental problems in mathematics, which is connected to many problems in mathematics and physics. Euler [16] revealed a series written as an infinite product over the prime numbers p. Riemann expanded this function by the tools of complex analysis [54]. Hardy [23] proved that infinitely many zeros lie on the critical line, but which is much weaker than the Riemann's conjecture, requiring all the nontrivial zeros to be on the critical line. Borwein et al. [7] gave a mathematical introduction to the history of the Riemann hypothesis and its equivalent statements.

Several approaches have attempted to solve the problem. 1) To find the nontrivial zeros of the Riemann zeta function $\zeta(s)$ by computing projects. If anyone of the nontrivial zeros were located off the critical line, one would disprove the Riemann's conjecture. But up to date, no such a nontrivial zero has been found. Although 100 billions nontrivial zeros have been found to lie on the critical line [20], it is too far for proving/disproving the Riemann's conjecture. 2) To determine how many percentage of the nontrivial zeros lie on the critical line. Levinson [37] showed that at least one-third of the nontrivial zeros are on the critical line which was later incrementally improved to two-fifths [10]. Clearly, it is extremely difficult to prove the Riemann hypothesis along this path. 3) To give the zero-density estimates for the Riemann zeta function $\zeta(s)$. Let $N(\sigma, T)$ be the number of zeroes of $\zeta(s)$ in the rectangle $\Re(s) \geq \sigma$ and $|\Im(s)| \leq T$. Ingham [28] proved the bound $N(\sigma, T) \leq T^{\frac{3(1-\sigma)}{2-\sigma}+o(1)}$. Recently, Guth and Maynard [22] deduced a zero density estimate $N(\sigma, T) \ll T^{\frac{15(1-\sigma)}{3+5\sigma}+o(1)}$ and combing with Ingham's estimate when $\sigma \leq 7/10$ they obtained $N(\sigma, T) \leq T^{\frac{30(1-\sigma)}{13}+o(1)}$. However, this new advance is still far from the desired solution. 4) To connect the Riemann zeta function $\zeta(s)$ with physical systems. The main obstacle of the path is to find an appropriate model to satisfy the condition that the zero distribution of the partition function is equivalent to the distribution of the nontrivial zeros of the Riemann zeta function $\zeta(s)$.

On the other hand, the Ising model is one of the most fundamental models in physics [29], which is associated with many problems in physics, mathematics and computer science. The study of the mathematical structures of the Ising model is helpful for a better understanding of other mathematical problems [52]. The present author has solved two fundamental problems related with Ising models: the exact solution of the ferromagnetic three-dimensional (3D) Ising model [64,71,73,74,76,80.81] and the computational complexity of the spin-glass 3D Ising model [75,78,79,82]. In the first case, the exact solution has been supported by experimental data [24,42,57,72] and Monte Carlo simulations [38,39]. In the second case, the lower bound for the computational complexity of the spin-glass 3D Ising model (and also the Boolean satisfiability problem) has been recognized by several groups of computer scientists worldwide [30,41,50,56,69]. Because the Riemann hypothesis is closely related with the statistical physics [59,67], I believe that the Ising model may play an important role for solving the Riemann hypothesis. The key is to set up a reliable Ising model, which has random distributions of energy eigenvalues, eigenvectors and zeros of its partition function, and its zero distribution is the same as that of the Riemann zeta function $\zeta(s)$.

The aim of this work is to prove the equivalence between the zero distributions of the Riemann zeta function $\zeta(s)$ and a two-dimensional (2D) Ising model with a mixture of ferromagnetic and randomly distributed competing ferromagnetic/antiferromagnetic interactions. In section 2, we review briefly the characteristics of the Riemann hypothesis and its connections to physics. In section 3, we build a 2D Ising model with ferromagnetic interactions and randomly distributed competing ferromagnetic/antiferromagnetic interactions along two crystallographic directions respectively. We prove that all energy eigenvalues of this 2D Ising model are real and randomly distributed as the $L(s, \chi_k)$ function as well as the Riemann zeta function $\zeta(s)$ and that its eigenvectors are constructed by the eigenvectors of the 1D Ising model with phases related to the Riemann zeta function $\zeta(s)$, via the relation $\omega_1(\gamma_{2j})$ between the angle $\omega_1$ and the energy eigenvalues $\gamma_{2j}$. We also prove that all the zeros of the partition function of the 2D Ising model lie on a unit circle in the parametric space of complex temperature (i.e. Fisher zeros), which can be mapped to the zero distribution of the $L(s, \chi_k)$ function and also the Riemann zeta function $\zeta(s)$ in the critical line. Namely, we have proven the closure of the nontrivial zero distribution of the $L(s, \chi_k)$ function (including the Riemann zeta function $\zeta(s)$).

## 2. Notation and characteristics of the Riemann hypothesis and its connections to physics

*Notation.*

$p$, $p_1, p_2, p_r, p_n$: a prime number.

$q$, $X$: an even number.

$n$, $a$: an integer number.

$x$: a number.

$L(s, \chi_k)$: the Dirichlet function.

$\zeta(s)$: the Riemann zeta function.

$\mu(n)$: the Möbius function.

$\Lambda(n)$: the von Mangoldt function.

$P(\tilde{J}_2)$: the Gaussian function.

$\chi_k(n)$: a primitive Dirichlet character.

$k$, $h$: an integer modulo of the Dirichlet character.

$\boldsymbol{R}$: an operator.

$\boldsymbol{I}$: the unit matrix.

$\boldsymbol{H}$: a self-adjoint operator interpreted as a Hamiltonian of a physical system.

$\rho_k$: the nontrivial zero of the Dirichlet function.

$\gamma_k$, $E$, $E_n$: the imaginary part of the Dirichlet function.

$s_{\mathrm{i,j}}$: the Ising spin.

i, j, t, n, m: the lattice site.

$J$, $J_0$, $J_1$, $J_2$, $\tilde{J}_2$: the interaction between spins.

$K$, $K_1$, $K_1^*$, $\widetilde{K}_2$: the interaction variable with respect to temperature.

$\mu'(i, j)$: a function imitating the Möbius function.

$k_B$: the Boltzmann constant.

$T$: temperature.

Z, $\bar{Z}_\alpha$: the partition function.

α, $R$: the replica of the spin-glass system.

**V**, $\mathbf{V_1}$, $\mathbf{V_2}$: the transfer matrix.

$\boldsymbol{U}$: the boundary factor in the transfer matrix.

$\Gamma_{2j}$, $C_j$, ${s'}_j$: the matrix in the spinor representation.

$\sigma_1$, $\sigma_2$, $\sigma_3$, $C$, $\mathrm{s}'$, $\mathrm{s}''$: the Pauli matrix.

$\boldsymbol{R_0^-}$: the rotation transformation matrix.

$\boldsymbol{\alpha}_{2j}$, $\boldsymbol{a}$, $\boldsymbol{b}$, $\boldsymbol{b}^*$: a matrix in the rotation transformation matrix.

$\gamma_{2j}$: an energy eigenvalue.

$\omega_1$, $\delta'$, $\delta_t^*$: an angle in a hyperbolic triangle in the 2D Poincaré disk model.

$E_j$: an energy level.

$u_{2j}$, $v_{2j}$: the 2n·l·o-normalized eigenvectors.

$v_\theta$, $v_c$: the locus of the unit circle.

$x$, $x_0$, $x_1$, $u$, $v$, $s$: a complex variable.

## 2.1 Riemann hypothesis

Using the fundamental theorem of arithmetic, Euler [16] revealed a series written as an infinite product over the prime numbers, p,

$$\zeta(\mathrm{k}) = \sum_{n=1}^{\infty} \frac{1}{n^k} = \prod_p \left(1 - \frac{1}{p^k}\right)^{-1} \tag{1}$$

Riemann applied the tools of complex analysis to this function and proved that the function defined by the infinite summation [54],

$$\zeta(s) = \sum_{n=1}^{\infty} \frac{1}{n^s} \tag{2}$$

can be analytically continued over the complex s plane, except for s = 1, which is called the Riemann zeta function. Here, s denotes a complex number $s = \sigma + it$, where $\sigma$ and t are real numbers and i is the imaginary unit. Riemann derived a functional equation, containing the $\zeta(s)$ function, which is valid for all complex s and exhibits mirror symmetry around the $\sigma = 1/2$ vertical line, called the critical line, such that

$$\pi^{-s/2}\Gamma\left(\frac{s}{2}\right)\zeta(s) = \pi^{-(1-s)/2}\Gamma\left(\frac{1-s}{2}\right)\zeta(1-s)$$

for $s \in \mathbb{C}\{0,1\}$

(3)

**Riemann's Hypothesis [54]:** All nontrivial zeros of the function $\zeta(s)$ have the form $\rho = 1/2 + it$, where t is a real number. In other words, all nontrivial zeros lie on the critical line.

The prime number theorem states that the prime-counting function $\pi(x)$ asymptotically behaves as the logarithmic integral Li($x$) [58,66]:

$$\pi(x) \sim \mathrm{Li}(x) \coloneqq \int_{2}^{x} \frac{du}{\ln(u)} \approx \frac{x}{\ln(x)} \tag{4}$$

Then $\pi(x)$ is given by the series via so called Möbius inversion formula [66]:

$$\pi(x) = \sum_{n\geq 1} \frac{\mu(n)}{n} J\left(x^{1/n}\right) \tag{5}$$

where the sum is in fact finite because it stops at such $N$ that $x^{1/N} > 2 > x^{1/(N+1)}$ and $J(x)$ is defined as

$$J(x) = \pi(x) + \frac{1}{2}\pi\left(x^{1/2}\right) + \frac{1}{3}\pi\left(x^{1/3}\right) + \cdots \tag{6}$$

and $\mu(n)$ is the Möbius function

$$\mu(n) = \begin{cases} 1, & for\ \ n = 1 \\ 0, & when\ n\ is\ divisable\ by\ a\ square\ of\ a\ prime\ p\colon\ p^2\ |n \\ (-1)^r, & where\ \ n = p_1 p_2 \dots p_r \end{cases} \tag{7}$$

The distribution of prime numbers among the integers is a fundamental problem of number theory (see, e.g. [48]). It is connected closely to the properties of Dirichlet L functions (including the Riemann zeta function), defined via [7,63]

$$L(s, \chi_k) \coloneqq \sum_{n=1}^{\infty} \frac{\chi_k(n)}{n^s} \tag{8}$$

for $\mathrm{Re}(s) > 1$ and by analytic continuation elsewhere, where $\chi_k(n)$ is a primitive Dirichlet character: $\chi_k(n)$ is a number-theoretic character modulo k, defined by the properties $\chi_k(mn) = \chi_k(m)\chi_k(n)$, $\chi_k(n) = \chi_k(n+k)$, $\chi_k(1) = 1$ for all integers, and $\chi_k(n) = 0$ for $(k, n) \neq 1$. The Riemann zeta function is given by $\zeta(s) \coloneqq L(s, \chi_1)$.

The definition of $\mu(n)$ stems from the formula (2) and the Dirichlet series for the reciprocal of the zeta function $\zeta(s)$ [7,11,33,66]:

$$\frac{1}{\zeta(s)} = \prod_{p=2,3,5,7\ldots}\left(1-\frac{1}{p^s}\right) = \sum_{n=1}^{\infty}\frac{\mu(n)}{n^s} = s\int_1^{\infty}\frac{M(x)}{x^{s+1}}dx \qquad (Re(s)>1)$$

(9)

Then the number of primes up to $x$ is obtained [66]:

$$\pi(x) = \sum_{n=1}^{N}\frac{\mu(n)}{n}\left(Li\left(x^{1/n}\right) - \sum_{\rho} Li\left(x^{\rho/n}\right)\right)$$

(10)

since $J(x) = Li(x) - \sum_{\rho} Li(x^{\rho})$. Here

$$M(x) \coloneqq \sum_{n\leq x}\mu(n)$$

(11)

Thus a Mertens type estimate [33]:

$$M(x) = O_{\varepsilon}\left(x^{\frac{1}{2}+\varepsilon}\right)$$

(12)

for all $\varepsilon > 0$ would imply convergence of (9) in the half plane Re($s$) > 1/2 and thus the Generalized Riemann Hypothesis.

**Generalized Riemann Hypothesis [7,9]**: All nontrivial zeros of the Dirichlet function $L(s,\chi_k)$ have real part equal to $\frac{1}{2}$.

According to the generalized Riemann hypothesis [63], any zero of the Dirichlet function $L(s,\chi_k)$ with $0 < Re(s) < 1$ is on the critical line $Re(s) = \frac{1}{2}$; these are the nontrivial zeros, which we write as $\rho_k = \frac{1}{2} + i\gamma_k$ The generalized Riemann hypothesis implies that each $\gamma_k$ is real, and this in turn implies that the number of primes less than x in the arithmetic progression a; a+ k; a + 2k; . . . (with a less than and coprime to k) is, in the limit of large x,

$$\pi_{a,k}(x) = \frac{1}{\varphi(k)} Li(x) + O(x^{\frac{1}{2}+\epsilon})$$

(13)

or all $\epsilon > 0$, where $\varphi(k)$ is the number of integers less than and coprime to k (the Euler totient function), and Li(x) is the logarithmic integral function. The exponent of x in the error term increases to $\frac{1}{2} + \max(\mathrm{Im}\gamma_k) + \epsilon$ if the generalized Riemann hypothesis is false.

**2.2 Connections to physics**

It is noticed the fruitful and diverse area of extensions of the Riemann zeta function $\zeta(s)$ which also occur throughout physics [58,66], such as, classical mechanics (for instance, billiards), quantum mechanics (for instance, quantum billiards, scattering state models and bound state models), nuclear physics (for instance, random matrix theory), condensed matter physics (for instance, solid built up by cations and anions) and statistical physics.

It is an old idea, now generally known as the Hilbert-Pólya conjecture (see [13] for a historical review):

**Hilbert-Pólya Conjecture [58]:** The nontrivial zeros of the Dirichlet function $L(s, \chi_k)$ (including the Riemann function ζ(s)) can be the spectrum of an operator, $\mathbf{R} = \frac{1}{2}\boldsymbol{I} + i\boldsymbol{H}$ where $\boldsymbol{I}$ is the unit matrix, $\boldsymbol{H}$ is self-adjoint operator interpreted as a Hamiltonian of a physical system.

Each Dirichlet function $L(s, \chi_k)$ would have a different Hamiltonian $\boldsymbol{H}$. Furthermore, $L(\frac{1}{2} + iE, \chi_k)$ is conjectured to be proportional to the spectral determinant $\det(E-H)$; since the eigenvalues of a Hermitian operator must be real, the Hilbert-Pólya conjecture implies the generalized Riemann hypothesis [63]. A large body of analytic and numerical work strongly supports the Montgomery-Odlyzko law (see, e.g., [11,46]), which states that the statistical distribution of the imaginary part $\gamma_k$ for each Dirichlet function $\mathrm{L}(s,\chi_k)$ is the same as the Wigner-Dyson distribution of the eigenvalues of large Hermitian matrices with real diagonal entries and complex off-diagonal entries, each selected from a Gaussian distribution; this is the Gaussian unitary ensemble [32,44].

In what follows, we shall introduce briefly some examples for possible physical systems.

We can construct the physical system with the Riemann zeta function $\zeta(\mathrm{s})$ as a partition function. For instance, the problem of construction of a one-dimensional (1D) Hamiltonian whose spectrum coincides with the set of primes was considered in [49,59,60], see also review [55,66]. The comparison of the number of $\zeta(\mathrm{s})$ zeros and the number of energy eigenvalues below a threshold suggested that the physical system is quasi-one dimensional [59]. Some modification should lead to the Hamiltonian $\boldsymbol{H}$ having eigenstates $|p\rangle$ labeled by the prime numbers $p$ with eigenvalues $E_p=\mathcal{E}\ln(\mathrm{p})$ where $\mathcal{E}$ is some constant with dimension of energy. The $n$ particle state can be decomposed into the states $|p\rangle$ using the factorization theorem. The energy of the state $|n\rangle$ is equal to $E_n=\mathcal{E}\sum_{i=1}^{k}\alpha_i\ln(p_i)=\mathcal{E}\ln(\mathrm{n})$. Then the partition function $Z$ is given by the Riemann zeta function $\zeta(s)$ [66]:

$$Z(T) = \sum_{n=1}^{\infty} exp\left(-\frac{E_n}{k_B T}\right) = \sum_{n=1}^{\infty} exp\left(-\frac{\mathcal{E}\ln(n)}{k_B T}\right) = \sum_{n=1}^{\infty} \frac{1}{n^s} = \zeta(s)$$

(14)

with $s \equiv \mathcal{E}/k_B T$.

The Riemann zeta function occurs in numerous different branches of statistical physics, from Brownian motion to lattice gas models. The combinatorial problem arises in many different branches of mathematical physics, such as lattice animals in statistical physics [40,68], numerical analysis on combinatorial optimization [45]. The results in physics can have profound implications for mathematics in general and number theory. The interpretation of prime numbers or the Riemann zeta zeros as energy eigenvalues of particles appears not just in quantum mechanics but also in statistical mechanics. For instance, one may introduce two concepts: the Riemann gas, sometimes called the primon gas, and the Riemann liquid, although their definitions vary slightly. from this spectrum [58]

$$Z_B = \sum_{n=1}^{\infty} exp\left(-\frac{E_n}{k_B T}\right) = \sum_{n=1}^{\infty} \frac{1}{n^s} = \zeta(s)$$

(15)

where $s = \frac{\epsilon_0}{k_B T}$. Here the total energy of the system in the state $|n\rangle$ is $E_n = \epsilon_0 \ln(n)$. The partition function for the primon gas is thus the Riemann zeta function $\zeta(s)$ and hence the alternative nomenclature.

The particular interest of the present work is the statistical physics, and thus we shall focus on it. It is well known that the lattice gas models can be mapped into Ising models at the magnetic field [36,70]. Therefore, the conclusion above for the physical systems can be extended to be applicable for the partition function $Z$ of the Ising models. The Möbius function μ(n) can be treated as a q =3 Potts model or a spin-glass Ising model. If one defines the random motion through the Möbius function μ(n), i.e., if μ(n) = ±1, the particle moves up or down, and if μ(n) = 0, it does not move. For the q =3 Potts model [68], the spins have three states (up, down and empty). For the ferromagnetic Ising model (i.e., the q =2 Potts mode), the spins have two states (up and down). But for the spin-glass Ising model when the random interactions vary in [-J, J], an additional state with empty will appear effectively (see the next section for details). The functional equation (3) can be written in non-symmetrical form [66]:

$$2\Gamma(s)\cos\left(\frac{\pi}{2}s\right)\zeta(s) = (2\pi)^s\zeta(1-s) \tag{16}$$

In this form it is analogous to the Kramers–Wannier [34] duality relation for the partition function $Z(K)$ of the 2D Ising model with parameter $K$ expressed in units of $k_B T$ (i.e. $K = J/(k_B T)$) [66]

$$Z(K) = 2^N(coshK)^{2N}(tanhK)^N Z(K^*), \tag{17}$$

where $N$ denotes the number of spins and $K^*$ is related to $K$ via $e^{-2K^*} = tanhK$. Readers may refer also to [33] for an approach based on a statistical mechanics interpretation of the Riemann zeta function $\zeta(s)$.

To pursue this analogy between the Ising model and the Dirichlet function $L(s,\chi_k)$ (including the Riemann $\zeta(s)$ function), the key to solve the problem is to set up a suitable model to satisfy the following conditions: 1) All the discrete energy eigenvalues are real, which are randomly distributed the same way as the Dirichlet function $\mathrm{L}(s,\chi_k)$ (including the Riemann $\zeta(s)$ function); 2) All the zeros of the partition function of the model lie on a unit circle, which can be mapped into a critical line. In the next section, we shall establish a 2D model with randomly distributed competing interactions, derive the exact solution for the energy eigenvalues and the partition function of this model. We shall prove that this special model satisfies both the conditions and gives an unique solution for the nontrivial zeros of the Dirichlet function $\mathrm{L}(s,\chi_k)$ (including the Riemann $\zeta(s)$ function) in the critical line.

## 3. 2D Ising model with randomly distributed competing interactions

### 3.1 Hamiltonian and transfer matrices

**Definition 1.** A 2D Ising model with ferromagnetic interactions and randomly distributed competing ferromagnetic/antiferromagnetic interactions along two crystallographic directions respectively is defined as $M_{FI+SGI}^{2D}$. Here the subscript FI denotes the ferromagnetic Ising spins, while SGI denotes the spin-glass Ising spins with randomly distributed competing interactions.

**Theorem 1 (Equivalence Theorem).** The zero distribution of the partition function of the 2D Ising model $M_{FI+SGI}^{2D}$ is equivalent to the zero distribution of the Dirichlet function $\mathrm{L}(s,\chi_k)$ (including the Riemann zeta function $\zeta(s)$).

**Proof of Theorem 1.** The Hamiltonian of the 2D Ising model, $M_{FI+SGI}^{2D}$, is written as [29,31,52]:

$$\mathrm{H} = -\sum_{<i,j>}^{m,n} \left[ J_1 s_{i,j} s_{i+1,j} + \tilde{J}_2 s_{i,j} s_{i,j+1} \right] \tag{18}$$

Here every Ising spin takes two values +1 and -1 for spin up and spin down, respectively, which is located on a 2D lattice with the lattice size $N = mn$. The numbers (i, j), running from (1, 1) to ($m$, $n$), denote lattice points along two crystallographic directions. Only are the nearest neighboring interactions between spins located at a 2D lattice considered. The first interaction $J_1 > 0$ is ferromagnetic, while the second one $\tilde{J}_2$ is a randomly distributed competing ferromagnetic/antiferromagnetic interaction as in a spin-glass system [15,47,53]. The distribution $\mathrm{P}(\tilde{J}_2)$ of values $\tilde{J}_2$ is taken to be a Gaussian function with a non-zero mean $J_0 \neq 0$ and a variance $\hat{J}_2^2$. Namely, the interaction $\tilde{J}_2$ varies randomly with different signs in range of [-$J_2$, $J_2$] with $J_2 \geq 0$. So, the present model can be viewed as a 2D mixture model consisting of a 1D ferromagnetic Ising model and a 1D spin-glass Edwards-Anderson model [15]. Note that the interaction $J_2$ varies from zero to infinite, to fit with the unbounded Gaussian distribution. This model is the same as random layered frustration Ising models investigated in details in [25,26,43,67].

At first, we can use the Ising model to imitate the Möbius function μ(n). An Ising spin have two states (up and down), i.e., $s_i = \pm 1$. For the nearest neighboring interactions $s_i s_{i+1}$ between two spins, the combination of the spins states results in the two possibilities +1 and -1, which correspond to $s_i$ and $s_{i+1}$ have the same and different signs (alignments) respectively. That is:

$$s_{i,j}s_{i,j+1} = \begin{cases} +1, & if \;\; s_{i,j} = +1, s_{i,j+1} = +1; or \;\; s_{i,j} = -1, s_{i,j+1} = -1. \\ -1, & if \;\; s_{i,j} = +1, s_{i,j+1} = -1; or \;\; s_{i,j} = -1, s_{i,j+1} = +1. \end{cases}$$

(19)

The Möbius function $\mu$(n) can be treated as a $q$ =3 Potts model with three spin states (up, down and empty) [68]. The three states spin-1 Ising Hamiltonian was mapped onto the two state spin-1/2 Ising Hamiltonian by using the Griffith's variable transformation [21,83]. Thus, the Griffith's variable transformation ensures the relation between the Möbius function and the spin-1/2 Ising model via the mapping between the spin-1/2 and spin-1 Ising models. Furthermore, the three states can be realized in a spin-glass spin-1/2 Ising model with two spin states (up and down) and an additional state (empty) induced by random interactions, which can be described as the following procedure.

When the random interactions $\tilde{J}_2$ are introduced, which vary in $[-J_2, J_2]$, an additional state with empty will appear effectively when $\tilde{J}_2 = 0$. All the characters together give a function

$$\mu'(\mathrm{i,j}) = \tilde{J}_2 s_{\mathrm{i,j}} s_{\mathrm{i,j+1}}$$

$$= \begin{cases} \tilde{J}_2, & for \;\; s_{\mathrm{i,j}} \;\; and \;\; s_{\mathrm{i,j+1}} \;\; have \;\; the \;\; same \;\; sign \;\; and \;\; \tilde{J}_2 \neq 0. \\ 0, & when \;\; \tilde{J}_2 = 0 \\ -\tilde{J}_2, & for \;\; s_{\mathrm{i,j}} \;\; and \;\; s_{\mathrm{i,j+1}} \;\; have \;\; the \;\; different \;\; sign \;\; and \;\; \tilde{J}_2 \neq 0. \end{cases}$$

(20)

The function $\mu'(\mathrm{i,j})$ can be normalized so that it can be used to imitate the Möbius function μ(n) (Eq. (7)), Dirichlet $\mathrm{L}(s, \chi_k)$ functions (Eq. (8)) and thus the Riemann function $\zeta(\mathrm{s})$ (Eq. (2)), by adjusting the distribution $\mathrm{P}(\tilde{J}_2)$. Therefore, we can use the competing 2D Ising model to imitate the Möbius function $\mu$(n).

It is noticed that the Möbius function and Dirichlet characters depend on a single integer n and all their properties follow from the arithmetic features of this $n$ with respect to the modulus $q$. The function $\mu'(\mathrm{i},\mathrm{j})$ depends on two indices i and j, used for the 2D lattice. The two indices (i, j) run from (1, 1) to (m, n), can be represented by a single integer $\mathrm{J} = [(j-1)m+i]$ = 1,2,…,m,m+1, …, 2m, 2m+1, …, m(n-1), m(n-1)+1,…, mn. Furthermore, the products in the transfer matrices $\mathbf{V}_2$ and $\mathbf{V}_1$ (Eqs. (11) and (12)) run over j from 1 to n, after performing the periodic condition, which may set a one-to-one correspondence with the Möbius function and Dirichlet characters. It is understood that it is not very important to determine exactly the location of all the nontrivial zeros of the Dirichlet function $\mathrm{L}(s,\chi_k)$ (including the Riemann zeta function $\zeta(s)$), but imitate its trend for the zero distribution, in particular, in the large number limit (corresponding to the high energy limit). It is extremely important to prove the closure of the nontrivial zero distribution, while excluding the possibility that the nontrivial zeros lie off the critical line. The zero distribution in the large number limit is the most important, but merely uncertain factor for proving its closure. By utilizing the properties of Ising model together with randomly distributed competing interactions, we can realize this purpose by ergodic of all replicas [47,53].

By pretending that the imaginary parts $E_m$ of the Riemann zeros are eigenvalues of a quantum Hamiltonian whose corresponding classical trajectories are chaotic and without time-reversal symmetry, Berry [2] obtained the mean square difference between the actual and average numbers of Riemann zeros near the $x$-th zero in an interval. Berry and Keating [3] gave the Riemann zeros and eigenvalue asymptotics. Forr12.ester et al. [18] studied the distributions of energy eigenvalues of the Gaussian unitary ensemble and the zeros of the Riemann zeta function ζ(s), described by nonlinear equations. Bogomolny [5] discussed statistical properties of quantum eigenvalues for chaotic systems based on semiclassical trace formulas, and considered the statistics of the zeros of the Riemann zeta function $\zeta(\mathrm{s})$. Conrey and Snaith [12] investigated correlations of eigenvalues of a unitary matrix (and its conjugate transpose) and Riemann zeros. Languasco et al. [35] obtained a quantitative version of the Goldston-Montgomery theorem about the equivalence between the asymptotic behaviors of the mean-square of primes in short intervals, and of the pair correlation function of the nontrivial zeros of the Riemann zeta function $\zeta(\mathrm{s})$. It is commonly accepted that the statistical distribution of the imaginary part $\gamma_k$ of each Dirichlet function $L(s, \chi_k)$ is the same as the Wigner-Dyson distribution of the eigenvalues of the Gaussian unitary ensemble [14,46,65]. Because of the randomness of the interaction $\tilde{J}_2$ $(\tilde{K}_2)$ with the Gaussian distribution, the transfer matrix has the character of a random unitary matrix [14,65], and the present system behaves as the Gaussian unitary ensemble [32,44]. It is noticed that the distribution of the eigenvalues of the random matrices possess the universality as the order of the random matrices approaches infinite, which does not reply on the distribution of particular Hamiltonian of a physical system. Such a universality can be applicable for describing the distribution of energy levels of complex systems, from disorder mediums, neural networks to quantum chaos [58,66]. For more details of the relation between the Ising model, the random matrix and the Riemann zeta function ζ(s), readers refer also to [1,6,19,27,51].

The replica method has been employed for treating the random systems, like the spin-glass Edwards-Anderson model [15], which rests on the use of the exact relation $[lnZ\{x\}]_{av} = \lim_{R\to 0}\frac{1}{R}([Z^R\{x\}]_{av} - 1) = \lim_{R\to 0}\frac{\partial}{\partial R}[Z^R\{x\}]_{av}$ [4]. For positive integer $R$, one can express $Z^R\{x\}$ in terms of $R$ identical replicas of the system. The partition function Z of the 2D Ising model $M_{FI+SGI}^{2D}$ can be calculated from the product of the partition functions $\bar{Z}_\alpha$ for all fixed replicas (α = 1,2,..,R), $Z = \prod_{\alpha=1}^{R}\bar{Z}_\alpha$ (refer to Eq. (3.25) of [4]). Here α denotes the α-th replica. The primon gas system with the Riemann zeta function $\zeta(s)$ as a partition function can be constructed by a 1D Hamiltonian whose spectrum coincides with the set of primes (see Eq. (56) in [58] and Eq. (55) in [66]). Similarly, the partition function $\bar{Z}_\alpha$ of the 2D Ising model $M_{FI+SGI}^{2D}$ for each replica can be written as:

$$\bar{Z}_\alpha = \sum_{n=1}^{\infty} exp\left(-\frac{E_n}{k_B T}\right) = \sum_{n=1}^{\infty}\frac{1}{n^s} = \zeta(s) \tag{21}$$

with $s \equiv \mathcal{E}/k_B T$ and $E_n = \mathcal{E}\ln(n)$. This equivalence is validated for the Gaussian unitary ensembles including the present model as the order of the random matrices approaches infinite, corresponding to the thermodynamical limit. It ensures that such an equivalence is held for all the primes and all the eigenvalues, which certainly cover the large number limit of the zero distribution (*i.e,* the high energy limit of the energy levels). Clearly, we can see from Eq. (21) that $\bar{Z}_\alpha = 0 \Leftrightarrow \zeta(s) = 0$. Namely, the zeros of the partition function $\bar{Z}_\alpha$ of the 2D Ising model $M_{FI+SGI}^{2D}$ for each replica are equivalent to the zeros of the Riemann zeta function $\zeta(s)$. Because the change in the Hamiltonian $\boldsymbol{H}$ corresponds to the change in the Dirichlet function $\mathrm{L}(s,\chi_k)$, scanning over all the replicas would correspond to the change in the Dirichlet character $\chi_k$ of the Dirichlet function $\mathrm{L}(s,\chi_k)$. As a consequence, the zeros of the partition function Z of the 2D Ising model $M_{FI+SGI}^{2D}$ are equivalent to the nontrivial zeros of the Dirichlet function $\mathrm{L}(s,\chi_k)$.

Second, it is necessary to construct a 2D Ising model. No phase transition occurs in the 1D Ising models at the absence of a magnetic field. For the 2D ferromagnetic Ising model, all the interactions are ferromagnetic and there is a phase transition, but no randomness. For the 2D spin-glass Ising model, all the interactions are randomly distributed and no phase transition occurs [62], since the lower critical dimensionality for spin glass is about 2.5. The unique 2D Ising model is what we constructed here, $M_{FI+SGI}^{2D}$, with a mixture of ferromagnetic and randomly distributed competing interactions along the two crystallographic directions respectively. The present model $M_{FI+SGI}^{2D}$ in a certain sense satisfies the requirement for a quasi-one dimensional physical system [59], since it is not a full 2D ferromagnetic model. There might exist other models, for instance, a 2D Ising model $M_{AFI+SGI}^{2D}$ with antiferromagnetic interactions and spin-glass interactions, *etc.*, which can be also utilized to realize the desired solution for the Dirichlet function $\mathrm{L}(s,\chi_k)$ as well as the Riemann zeta function $\zeta(s)$, but not the focus of the present work.

For simplicity, we apply the cylindrical crystal model preferred by Onsager [52] and Kaufman [31], in which we wrap our crystal on cylinders (since there is a translational invariance along the direction i due to the ferromagnetic interactions). The partition function of the 2D Ising model, $M^{2D}_{FI+SGI}$, at the zero magnetic field is expressed as follows

$$Z = (2sinh2K_1)^{\frac{n}{2}} \cdot \mathrm{trace}(V)^m \equiv (2sinh2K_1)^{\frac{n}{2}} \cdot \sum_{i=1}^{2^n} \lambda_i^m \quad (22)$$

with the transfer matrix **V** = **V₂V₁** as:

$$\boldsymbol{V_2} = \prod_{j=1}^{n-1} exp\left[-i\widetilde{K}_2\Gamma_{2j}\Gamma_{2j+1}\right] exp\left[i\widetilde{K}_2\Gamma_1\Gamma_{2n}U\right] = \prod_{j=1}^{n-1} \exp\left[\widetilde{K}_2 s'_j s'_{j+1}\right] exp\left[\widetilde{K}_2 s'_1 s'_n U\right] \quad (23)$$

$$\boldsymbol{V_1} = \prod_{j=1}^{n} \exp\left[iK_1^*\Gamma_{2j-1}\Gamma_{2j}\right] = \prod_{j=1}^{n} \exp\left[K_1^* \cdot C_j\right] \quad (24)$$

Note that the products in the transfer matrices **V₂** and **V₁** (Eqs. (11) and (12)) run over j from 1 to n, after performing the periodic condition. It is convenient to introduce variables $K_1 \equiv J_1/(k_B\mathrm{T})$ and $\widetilde{K}_2 \equiv \tilde{J}_2/(k_B\mathrm{T})$ instead of interactions $J_1$ and $\tilde{J}_2$ along the two crystallographic directions. Because of the randomness of the interaction $\tilde{J}_2$ ($\widetilde{K}_2$), the transfer matrix $\boldsymbol{V_2}$ has the character of a random matrix [61,65], and behaves as the Gaussian unitary ensemble mentioned above [32,44]. For more details of the relation between the random matrix and the Riemann zeta function ζ(s), readers refer also to [6,19,27]. Here, the Kramers-Wannier relation gives the definition of $K_1^*$ in its dual lattice [34]:

$$K_1^* = -\frac{1}{2} ln(tanhK_1) \quad (25)$$

The Kramers–Wannier [34] duality is analogous to the Riemann zeta function ζ(s) (see Eqs.(16) and (17) for comparison, and also [33] for details).

We define the matrices $C_j$ and $s'_j$ as follows:

$$C_j = \mathrm{I} \otimes \mathrm{I} \otimes \ldots \otimes \mathrm{I} \otimes \mathrm{C} \otimes \mathrm{I} \otimes \ldots \otimes \mathrm{I} \tag{26}$$

and

$$s'_j = \mathrm{I} \otimes \mathrm{I} \otimes \ldots \otimes \mathrm{I} \otimes \mathrm{s}' \otimes \mathrm{I} \otimes \ldots \otimes \mathrm{I} \tag{27}$$

Following the Onsager–Kaufman–Zhang notation [31,52,71], we have: $s'' = \begin{bmatrix} 0 & -1 \\ 1 & 0 \end{bmatrix}$ $(= -i\sigma_2)$, $s' = \begin{bmatrix} 1 & 0 \\ 0 & -1 \end{bmatrix}$ $(= \sigma_3)$, $C = \begin{bmatrix} 0 & 1 \\ 1 & 0 \end{bmatrix}$ $(= \sigma_1)$, where $\sigma_j$ ($j$ = 1,2,3) are Pauli matrices, while I is the unit matrix. It is remarked that from the reality of the Ising Hamiltonian (Eq. (6)) the transfer matrices $\boldsymbol{V_2}$ and $\boldsymbol{V_1}$ are real also. However, the imaginary number emerges in $\boldsymbol{V_2}$ and $\boldsymbol{V_1}$ as written in the Clifford algebraic representation, since the $\Gamma_{2j}$ matrix is complex-valued, which consists of a complex-valued Pauli matrix $\sigma_2$ The product of two matrices $\Gamma_{2j}\Gamma_{2j+1}$ in $\boldsymbol{V_2}$ or $\Gamma_{2j-1}\Gamma_{2j}$ in $\boldsymbol{V_1}$ corresponds to a rotation angle $\theta$ for $e^{\pm i\frac{\theta}{2}}$ [31], which is complex-valued. The imaginary number appearing in the transfer matrices $\boldsymbol{V_2}$ and $\boldsymbol{V_1}$ results in a rotation angle γ for $e^{\pm\frac{\gamma}{2}}$ [31], and accordingly the hyperbolic functions appear in the rotation matrices and the eigenvalues (see Eqs. (28)-(33)).

In the transfer matrices, the boundary factor $U = C_1 C_2 \ldots C_n$ as in Kaufman's paper [31] can be viewed as the projection operator $P \equiv \frac{1}{2}(1 \pm U)$, splitting the Hilbert space into two subspaces, which does not alter the character of real eigenvalues. In the thermodynamic limit, the surface to volume ration vanishes for an infinite system so that the boundary factor can be neglected [74].

Note that the products in the transfer matrices $\boldsymbol{V_2}$ and $\boldsymbol{V_1}$ (Eqs. (23) and (24)) run over j from 1 to n, after performing the periodic condition along one crystallographic direction. For a 2D lattice, the planar rotations $K_1$ and $\widetilde{K}_2$ in the spinor representation can be transformed into the rotation representation. Therefore, the eigenvalues of the partition function can be calculated by the planar rotations in the rotation representation. Following the Onsager-Kaufman procedure [31,52], we have the first (and last) product representing the rotation:

$$\begin{bmatrix} cosh K_1^* & isinh K_1^* & & & & \\ -isinh K_1^* & cosh K_1^* & & & & \\ & & cosh K_1^* & isinh K_1^* & & \\ & & -isinh K_1^* & cosh K_1^* & & \\ & & & & \cdot & \\ & & & & & \cdot \end{bmatrix} \tag{28}$$

The middle products have the form:

$$\begin{bmatrix} cosh 2\widetilde{K}_2 & & & & & -isinh 2\widetilde{K}_2 \\ & cosh 2\widetilde{K}_2 & isinh 2\widetilde{K}_2 & & & \\ & -isinh 2\widetilde{K}_2 & cosh 2\widetilde{K}_2 & & & \\ & & & \cdot & & \\ & & & & \cdot & \\ isinh 2\widetilde{K}_2 & & & & & cosh 2\widetilde{K}_2 \end{bmatrix} \tag{29}$$

Compared with those in Kaufman's procedure [31], only is difference in our procedure the interaction $\widetilde{K}_2$ is randomly distributed.

The rotation transformation matrix $\boldsymbol{R_0^-}$ could be written schematically as [31]:

$$\boldsymbol{R_0^-} = \begin{bmatrix} \boldsymbol{a} & \boldsymbol{b} & 0 & 0 & 0 & \cdots & 0 & \boldsymbol{b}^* \\ \boldsymbol{b}^* & \boldsymbol{a} & \boldsymbol{b} & 0 & 0 & \cdots & 0 & 0 \\ 0 & \boldsymbol{b}^* & \boldsymbol{a} & \boldsymbol{b} & 0 & \cdots & 0 & 0 \\ 0 & 0 & \boldsymbol{b}^* & \boldsymbol{a} & \boldsymbol{b} & \cdots & 0 & 0 \\ & & & & & & & \\ & & & & & & & \\ \boldsymbol{b} & 0 & 0 & 0 & 0 & \cdots & \boldsymbol{b}^* & \boldsymbol{a} \end{bmatrix} \tag{30}$$

where

$$\boldsymbol{a} = \begin{bmatrix} cosh2\widetilde{K}_2 \cdot cosh2K_1^* & -icosh2\widetilde{K}_2 \cdot sinh2K_1^* \\ icosh2\widetilde{K}_2 \cdot sinh2K_1^* & cosh2\widetilde{K}_2 \cdot cosh2K_1^* \end{bmatrix} \tag{31}$$

$$\boldsymbol{b} = \begin{bmatrix} -\frac{1}{2}sinh2\widetilde{K}_2 \cdot sinh2K_1^* & isinh2\widetilde{K}_2 \cdot sinh^2K_1^* \\ -isinh2\widetilde{K}_2 \cdot cosh^2K_1^* & -\frac{1}{2}sinh2\widetilde{K}_2 \cdot sinh2K_1^* \end{bmatrix} \tag{32}$$

$$\boldsymbol{b}^* = \begin{bmatrix} -\frac{1}{2}sinh2\widetilde{K}_2 \cdot sinh2K_1^* & isinh2\widetilde{K}_2 \cdot cosh^2K_1^* \\ -isinh2\widetilde{K}_2 \cdot sinh^2K_1^* & -\frac{1}{2}sinh2\widetilde{K}_2 \cdot sinh2K_1^* \end{bmatrix} \tag{33}$$

It is noticed that Dirichlet characters are, in general, complex-valued and, as the Mobius function, depend on an integer n. From the reality of the Ising Hamiltonian (Eq. (18),) the rotation matrices are unitary, which are complex-valued, since one of Pauli matrices is complex-valued. The complex property can be seen clearly from the imaginary off-diagonal elements in the rotation matrices (Eqs. (28)-(33)). Indeed, the transfer matrices (Eqs. (23) and (24)) and the rotation matrices (Eqs. (28)-(33)) possess the Wigner-Dyson distribution of the eigenvalues of large Hermitian matrices with real diagonal entries and complex off-diagonal entries, each selected from a Gaussian distribution being the Gaussian unitary ensemble [32,44]. As mentioned above, the distribution of the eigenvalues of the random matrices possess the universality as the order of the random matrices approaches infinite, which corresponds to the thermodynamical limit. The distribution of the eigenvalues of the present model is the same as the statistical distribution of the imaginary part $\gamma_k$'s for each L function. Therefore, we have shown that the 2D Ising model $M_{FI+SGI}^{2D}$ behaves as the Möbius function μ(n) and thus Riemann zeta function $\zeta(s)$. The zero distribution of the partition function of the 2D Ising model $M_{FI+SGI}^{2D}$ is equivalent to the zero distribution of the Dirichlet function $\mathrm{L}(s, \chi_k)$ (including the Riemann zeta function $\zeta(s)$).

□

If the Hilbert-Pólya conjecture is true, all the energy levels $E_n$ (thus the constant $\mathcal{E}$) in Eq. (21) will be real; vice versa. In the remainder of this article, we shall prove this point. It should be noticed that since s is a complex number and if the energy constant $\mathcal{E}$ is a real number, the temperature T must be a complex. For the physical significance of the complex temperature, readers refer to [77] for detailed discussion. Thus, the zeros of the partition function are distributed in a complex temperature plane. In what follows, we shall investigate in details the eigenvalues, the partition function and its zero distribution of the 2D Ising model $M_{FI+SGI}^{2D}$. It should be emphasized that from the real energy levels one can derive a critical line for the nontrivial zeros, while from the zeros of the partition function one can derive again a critical line. Because the partition function is calculated from the energy levels of the states of the system, the critical line derived from the two paths is unique.

### 3.2 Eigenvalues

**Theorem 2 (Real Eigenvalues Theorem).** All energy eigenvalues of the 2D Ising model $M_{FI+SGI}^{2D}$ are real, which are randomly distributed as the Riemann zeta function $\zeta(s)$.

**Proof of Theorem 2.** Following the procedure developed by Onsager and Kaufman [31,52], we can derive the eigenvalues for our present model $M_{FI+SGI}^{2D}$. The first and last lines in the rotation transformation matrix $\boldsymbol{R}_{\boldsymbol{0}}^{-}$ (Eq.(12)) correspond to the edge states with properties for N → ∞. It is clear that the edge matrix elements result in the same formulas as other elements in the rotation matrix, which indicates that as the system approach infinite the character of the energy eigenvalues does not alter. The 2n-eigenvalues of the rotation transformation matrix $\boldsymbol{R}_{\boldsymbol{0}}^{-}$ are the eigenvalues of the n 2-dimensional matrices $\boldsymbol{\alpha}_{2j} = \boldsymbol{a} + \epsilon^{2j} \cdot \boldsymbol{b} + \epsilon^{2(n-1)j} \cdot \boldsymbol{b}^{*}$ with $\epsilon \equiv e^{\frac{i\pi}{n}}$ [31]. The determinant of this rotation transformation matrix is +1. Its eigenvalues could be written as $\exp(\pm \gamma_{2j})$, and $\gamma_{2j}$ could be determined by:

$$\frac{1}{2}\text{trace}(\boldsymbol{\alpha}_{2j}) = \frac{1}{2}(e^{\gamma_{2j}} + e^{-\gamma_{2j}}) = cosh\gamma_{2j}$$

$$= cosh2K_1^{*}cosh2\widetilde{K}_2 - sinh2K_1^{*}sinh2\widetilde{K}_2cos\omega_1$$

(34)

where $\omega_1 = \frac{2j\pi}{n}$. In the 2D Ising case, $\gamma_{2j}$ is geometrically the third side of a hyperbolic triangle, represented in a 2D Poincaré disk model, whose other two sides, $2\widetilde{K}_2$ and $2K_1^*$. The angle between the two sides $2\widetilde{K}_2$ and $2K_1^*$ is determined by the angle $\omega$. The Hamiltonian $\boldsymbol{H}$ (Eq. (18)) of the 2D Ising model $M_{FI+SGI}^{2D}$ is hermitian (self-adjoint), and the transfer matrices **V** (Eqs. (23) and (24)) as well as the rotation transformation matrix $\boldsymbol{R_0^-}$ (Eq. (30)) are self-adjoint. Even in the thermodynamic limit (m $\to \infty$, n $\to \infty$), the physical properties of the system keeps the hermitian (self-adjoint) character to be the observable physical quantities. Therefore, all the energy eigenvalues of this model are real. Furthermore, because the interaction is randomly distributed, the energy eigenvalues are randomly distributed as the Riemann zeta function $\zeta(s)$. This proves that there exists and only exists a critical line $\mathrm{i}E_n$ (or $\mathrm{i}\gamma_{2j}$ as denoted in Eq. (34)) in the complex plane, and that no nontrivial zeros lie off the critical line. The conclusion is valid for every replica, and thus for all the replicas of the present model, corresponding to the Dirichlet function $L(s,\chi_k)$. .

□

In order to study the thermodynamic properties, we can use the replica method for the spin glass to obtain the energy eigenvalues $\bar{\gamma}_{2j,\alpha}$ for each replica α, in which the interaction $\widetilde{K}_2$ equals to the mean value $K_0 = J_0/(k_B T)$ according to the Gaussian distribution after averaging the disorder. Note that the actual and average numbers of Riemann zeros behave in the same tendency, since the mean interaction vale $J_0$ for the Gaussian distribution is located in between the actual values [-$J_2$, $J_2$].

### 3.3 Eigenvectors

**Theorem 3 (Hilbert-Pólya Space Theorem).** The eigenvectors of the 2D Ising model $M_{FI+SGI}^{2D}$ are constructed by the eigenvectors of the 1D Ising model with phases related to the Riemann zeta function $\zeta(s)$.

**Proof of Theorem 3.** The 2n-normalized eigenvectors of $\boldsymbol{R}_{\boldsymbol{0}}^{-}$ would behave as [31,52]:

$$u_{2j} \equiv \frac{1}{(2n)^{1/2}} \begin{bmatrix} exp\left(i\left(\omega_{2j}+\frac{1}{2}\delta_{2j}{}'\right)\right) \\ iexp\left(i\left(\omega_{2j}-\frac{1}{2}\delta_{2j}{}'\right)\right) \\ exp\left(i\left(\omega_{4j}+\frac{1}{2}\delta_{2j}{}'\right)\right) \\ iexp\left(i\left(\omega_{4j}-\frac{1}{2}\delta_{2j}{}'\right)\right) \\ \cdot \\ \cdot \\ \cdot \\ \cdot \\ \cdot \\ iexp\left(i\left(\omega_{2nj}-\frac{1}{2}\delta_{2j}{}'\right)\right) \end{bmatrix} \tag{35}$$

and

$$v_{2j} \equiv \frac{1}{(2n)^{1/2}} \begin{bmatrix} iexp\left(i\left(\omega_{2j} + \frac{1}{2}\delta_{2j}{}'\right)\right) \\ exp\left(i\left(\omega_{2j} - \frac{1}{2}\delta_{2j}{}'\right)\right) \\ iexp\left(i\left(\omega_{4j} + \frac{1}{2}\delta_{2j}{}'\right)\right) \\ exp\left(i\left(\omega_{4j} - \frac{1}{2}\delta_{2j}{}'\right)\right) \\ \vdots \\ \vdots \\ exp\left(i\left(\omega_{2nj} - \frac{1}{2}\delta_{2j}{}'\right)\right) \end{bmatrix}$$

(36)

The phase factor $exp(i\omega_{2j})$ is related with $cos\omega$ in the energy eigenvalues (Eq. (34)), while the phase factors $exp\left(\frac{i}{2}\delta_{2j}{}'\right)$ and $exp\left(-\frac{i}{2}\delta_{2j}{}'\right)$ form the normalized eigenvectors of a one-dimensional Ising spin chain [31]. From the eigenvalues $\gamma_{2j}$ (Eq. (34)), we can derive the following formulation for the angle $\omega_1$:

$$\omega_1(\gamma_{2j}) = arccos\left\{\frac{cosh2K_1^* cosh2\widetilde{K}_2 - cosh\gamma_{2j}}{sinh2K_1^* sinh2\widetilde{K}_2}\right\}$$

(37)

On the other hand, we have from Eqs. (33) and (34) of [58]:

$$\zeta\left(\frac{1}{2} - iE\right) = \frac{1}{\lambda^{\frac{1}{2}-iE}} = \lambda^{-\frac{1}{2}+iE} = e^{\ln\left(\lambda^{-\frac{1}{2}+iE}\right)} = e^{\left(-\frac{1}{2}+iE\right)ln\lambda}$$

(38)

Namely, we have:

$$i\gamma_{2j} = \frac{ln\left[\lambda^{\frac{1}{2}}\zeta\left(\frac{1}{2} - i\gamma_{2j}\right)\right]}{ln\lambda}$$

(39)

Therefore, the normalized eigenvectors of the 2D Ising model $M_{FI+SGI}^{2D}$ are constructed by the eigenvectors of the 1D Ising model with phases related to the Riemann zeta function $\zeta(s)$, via the relation $\omega_1(\gamma_{2j})$ between the angle $\omega_1$ and the energy eigenvalues $\gamma_{2j}$, which form an Hilbert-Pólya space [8].

□

Note again that for each replica α, we can use $\bar{\gamma}_{2j,\alpha}$ and $K_0$ to replace $\gamma_{2j}$ and $\widetilde{K}_2$, respectively, in the above discussion (see Eqs. (35)-(39)).

### 3.4 Zeros of the partition function

**Theorem 4 (Unit Circle - Critical Line Theorem):** All the zeros of the partition function of the 2D Ising model $M_{FI+SGI}^{2D}$ lie on a unit circle in the complex temperature plane, which can be mapped into a critical line.

**Proof of Theorem 4.** After mediating lnZ over disorder (i.e., $\overline{\ln Z}$), the partition function $\bar{Z}_\alpha$ of the 2D Ising model $M_{FI+SGI}^{2D}$ in a fixed replica α (α = 1,2,..,R) is represented as [31],

$$\bar{Z}_\alpha = \frac{1}{2}(2sinh2K_1)^{\frac{mn}{2}} \cdot \left\{ \prod_{j=1}^{n}\left(2cosh\frac{m}{2}\bar{\gamma}_{2j,\alpha}\right) + \prod_{j=1}^{n}\left(2sinh\frac{m}{2}\bar{\gamma}_{2j,\alpha}\right) + \prod_{j=1}^{n}\left(2cosh\frac{m}{2}\bar{\gamma}_{2j-1,\alpha}\right) + \prod_{j=1}^{n}\left(2sinh\frac{m}{2}\bar{\gamma}_{2j-1,\alpha}\right) \right\} \tag{40}$$

where the eigenvalues $\bar{\gamma}_{2j,\alpha}$ and $\bar{\gamma}_{2j-1,\alpha}$ are determined by Eq. (34) after averaging the disorder. Again, in these eigenvalues, one can investigate actual numbers in details and also average numbers. After averaging the disorder, the interaction $\widetilde{K}_2$ in the eigenvalues equals to the mean value $K_0$ of the Gaussian distribution.

Eq. (21) shows that the zeros of the partition function correspond to the zeros of the Riemann zeta function $\zeta(s)$. This can be generalized to be appropriate for the Dirichlet function $\mathrm{L}(s, \chi_k)$. Each L function would have a different $\boldsymbol{H}$, vice versa. It is understood that each replica for the 2D Ising model $M_{FI+SGI}^{2D}$ gives a different Gaussian distribution [47,53], changing the interaction $\tilde{J}_2$ ($\widetilde{K}_2$), thus altering the Hamiltonian $\boldsymbol{H}$. The change of the Hamiltonian $\boldsymbol{H}$ (and thus the partition function $\bar{Z}$) by scanning all replicas is equivalent to the change of the character $\chi_k$ all the way in the Dirichlet function $\mathrm{L}(s, \chi_k)$.

The Lee-Yang zeros of the canonical partition function are calculated at (real) temperature in the complex magnetic-field plane [36,70]. Lee and Yang proved that the zeros in the 2D Ising model with a magnetic field lie on the unit circle of the complex magnetic field plane [36]. The distribution of roots determines completely the equation of state, and in particular its behavior near the positive real axis prescribes the properties of the system in relation to phase transitions. On the other hand, Fisher showed that the zeros in the 2D Ising model at the absence of a magnetic field lie on the unit circle of the complex temperature plane [17]. In the Fisher's article [17], two unit circles appear in Figure 19.1, which correspond to the 2D ferromagnetic and antiferromagnetic Ising models respectively. The unit circle of the present model with the mixture of ferromagnetic Ising and randomly distributed Edwards-Anderson interactions lies also in the complex plane of temperature, which is consistent with the Fisher zeros for the 2D ferromagnetic Ising model. The zeros of the canonical partition function lie on a unit circle in the plane of the complex variable $\mathrm{v} = \mathrm{tanhK}$ [17], here for simplicity we set $K_0 = K_1 = K$. The center of the unit circle is located at the point (-1, 0), while it lies on the locus

$$v_\theta = -1 + \sqrt{2}e^{i\theta} \quad (0 \le \theta \le 2\pi) \tag{41}$$

When $\theta = 0 \;\; or \;\; 2\pi,\; v_c = \sqrt{2} - 1$ is the critical point for the present model, where a zero approaches the positive real axis in the thermodynamic limit, corresponding to a phase transition; when $\theta = \pi$, $\mathrm{v} = -(\sqrt{2} + 1)$ corresponds to non-physical complex temperature [17]. Note that the situation is just the same in the complex parametric plane of $\mathrm{x} = \exp(-2\mathrm{K})$, since the bilinear transformation $\mathrm{x} = \frac{1-v}{1+v}$ carries a circle into a circle. It should be noted that the complex variable should be normalized by a factor of $\sqrt{2}$ for the 2D Ising model to be $v' = \frac{\mathrm{tanhK}}{\sqrt{2}}$, so that the radius of the unit circle equals to unit.

For the general case of $K_1 \neq K_0$, the zeros of the canonical partition function lie on a unit circle in the plane of the complex variable $\mathrm{x} = (x_1 x_0)^{1/2} = \exp(-K_1 - K_0)$, or $\mathrm{v} = tanh\frac{1}{2}(K_1 + K_0)$. Note that these parameters have to be normalized also.

Next, we are interested in the critical point of the present 2D Ising model for the phase transition. With the mixture of ferromagnetic interactions along one direction and randomly distributed competing interactions along another direction, we can employ the periodic condition along the first direction, and thus the largest eigenvalue principle is applicable in the thermodynamic limit. It is seen clearly from the term of $\lambda_i^m$ in Eq. (22). In the thermodynamic limit, we have m → ∞ and n → ∞, the partition function $\bar{Z}_\alpha$ of the 2D Ising model $M_{FI+SGI}^{2D}$ in a fixed replica α after averaging the disorder is represented as [52],

$$N^{-1} ln\bar{Z}_\alpha = ln2 + \frac{1}{2(2\pi)^2}\int_{-\pi}^{\pi}\int_{-\pi}^{\pi} ln\{cosh2K_1 cosh2K_0 - sinh2K_1 cos\omega_1 - sinh2K_0 cos\omega'\} d\omega_1 d\omega' \quad (42)$$

The partition function (42) and the critical point of the present 2D Ising model are the same as the Onsager's exact solutions [52]. The critical point of the present model on the square lattice (if $K_0 = K_1$) is located at

$$x_c = e^{-2K_c} = \sqrt{2} - 1 \quad (43)$$

It is clear that the partition function $\bar{Z}_\alpha$ of the present 2D Ising model for every replica is the same as described in Eq. (42).

Therefore, all the zeros of the partition function $\bar{Z}_\alpha$ of the 2D Ising model $M_{FI+SGI}^{2D}$ in a fixed replica lie on a unit circle in the complex temperature plane. This conclusion is validated for every fixed replica. The partition function Z can be calculated from the product of the partition functions $\bar{Z}_\alpha$ for all fixed replicas (α = 1,2,..,R), $Z = \prod_{\alpha=1}^{R} \bar{Z}_\alpha$. Thus, all the zeros of the partition function Z of the 2D Ising model $M_{FI+SGI}^{2D}$ lie on a unit circle in the complex temperature plane.

The unit circle $/u/ = 1$ can be mapped into the critical line $\frac{1}{2} + it = s$ via the transformation $\mathrm{u} = \frac{s}{1-s} = \frac{\frac{1}{2}+it}{\frac{1}{2}-it} \rightarrow \mathrm{s}$. From all energy eigenvalues of the 2D Ising model $M_{FI+SGI}^{2D}$ are real, we have proven the critical line of $\mathrm{i}E_n$ (Theorem 2). It restricts the domain of definition for the mapping from the unit circle to the critical line. Thus, by devising appropriate spin system (e.g. the present Ising model) with $Z$(β, $z$) expressed by the Riemann zeta function $\zeta(s)$, the Lee–Yang theorem together with Fisher zeros can be used to locate all the possible zeros of the latter function, distributing on the critical line. This proves that there has and only has a critical line in the complex temperature plane and that no nontrivial zeros lie off the critical line. In another word, we have proven the closure of the zero distribution of the Dirichlet $\mathrm{L}(s, \chi_k)$ function (including the Riemann zeta function $\zeta(s)$).

□

Note that in the Lee-Yang Theorem for the unit circle [36], no assumptions were made about (1) the range of the interaction u, (2) the dimensionality of the lattice, (3) the size and structure of the lattice and (4) even the periodicity property of the lattice. The same is true for the Fisher's zeros [17]. Therefore, not only the present model $M_{FI+SGI}^{2D}$ but also other models (such as, $M_{AFI+SGI}^{2D}$, $M_{FI+SGI}^{3D}$, etc.) can be utilized to prove Theorem 4.

**Theorem 5.** The nontrivial zeros of the Dirichlet function $\mathrm{L}(s, \chi_k)$ (including the Riemann zeta function $\zeta(s)$ is the spectrum of an operator, $\mathbf{R} = \frac{1}{2}\boldsymbol{I} + i\boldsymbol{H}$ where $\boldsymbol{I}$ is the unit matrix, $\boldsymbol{H}$ is self-adjoint operator interpreted as the Hamiltonian of the 2D Ising model $M_{FI+SGI}^{2D}$

**Proof of Theorem 5.** Theorem 2 shows that the model $M_{FI+SGI}^{2D}$ is a suitable system for proving the Hilbert-Pólya conjecture [58]. All energy eigenvalues of the 2D Ising model $M_{FI+SGI}^{2D}$ are real, indicating that there is merely a critical line of $\mathrm{i}E_n$. Furthermore, according to Theorem 4, the unit circle for the zeros of the partition function of the 2D Ising model $M_{FI+SGI}^{2D}$ in the complex temperature plane also assures the existence of only a critical line. This has been proven to be true for every replica, and thus the conclusion is not only suitable for the Riemann zeta function $\zeta(s)$, but also for the Dirichlet function $\mathrm{L}(s, \chi_k)$. The critical line obtained in the two approaches is the same one, confirming the uniqueness of the critical line. Moreover, it is verified that no nontrivial zeros lie off the critical line. However, this ensures only the imaginary part of $\boldsymbol{R}$, while the real part needs to be fixed.

On the other hand, Hardy [23] proved that infinitely many zeros lie on the critical line. Up to date, 100 billions nontrivial zeros have been found to lie on the critical line with the real part as 1/2. The Hardy's result and the computation of the nontrivial zeros of the Riemann zeta function $\zeta(s)$ fixes already the real part of $\boldsymbol{R}$. If our critical line is located at $\sigma = 1/2$, it will be consistent with the Hardy's proof and the computation of the nontrivial zeros, else it will be contradictory with them. This excludes the possibility that the critical line is off $\sigma = 1/2$. Therefore, we have proven the Hilbert-Pólya conjecture.

□

## 4. Conclusion

In conclusion, we have proven five theorems: 1) The zero distribution of the partition function of the 2D Ising model $M_{FI+SGI}^{2D}$ is equivalent to the zero distribution of the Dirichlet function $\mathrm{L}(s,\chi_k)$ (including the Riemann zeta function $\zeta(s)$) (Equivalence Theorem). 2) All eigenvalues of the 2D Ising model $M_{FI+SGI}^{2D}$ are real, which are randomly distributed as the Dirichlet function $\mathrm{L}(s,\chi_k)$ and the Riemann zeta function $\zeta(s)$ (Real Eigenvalues Theorem). 3) The eigenvectors of the 2D Ising model $M_{FI+SGI}^{2D}$ are constructed by the eigenvectors of the 1D Ising model with phases related to the Riemann zeta function $\zeta(s)$ (Hilbert-Pólya Space Theorem). 4) All the zeros of the partition function of the 2D Ising model $M_{FI+SGI}^{2D}$ lie on a unit circle in the complex temperature plane, which can be mapped into a critical line (Unit Circle - Critical Line Theorem). 5) The closure of the nontrivial zero distribution of the $\mathrm{L}(s,\chi_k)$ function (including the Riemann zeta function $\zeta(s)$) has been proven. This work offers a novel thought to understand in-depth the generalized Riemann hypothesis and a path to upgrade it.

**Acknowledgements**

This work has been supported by the National Natural Science Foundation of China under grant numbers 52031014. The author is grateful to Fei Yang for understanding, encouragement, support and discussion.

**Declaration of competing interest:** The author declares no competing interests.

**References:**

1. N. Argaman, F. M. Dittes, E. Doron, P. Keating, A.Yu. Kitaev, M. Sieber, and U. Smilansky, Correlations in the actions of periodic orbits derived from quantum chaos, Phys. Rev. Lett. 71 (1993) 4326-4329.

2. M.V. Berry, Semiclassical formula for the number variance of the Riemann zeros, Nonlinearity 1 (1988), 399-407.

3. M.V. Berry, J.P. Keating, The Riemann zeros and eigenvalue asymptotics, SIAM Rev. 41 (1999) 236-266.

4. K. Binder, A.P. Young, Spin glasses: Experimental facts, theoretical concepts, and open questions, Rev. Mod. Phys. 58 (1986), 801-976.

5. E. Bogomolny, Spectral statistics and periodic orbits, Proc. Int. School of Physics 'Enrico Fermi' (Varenna) (1999), 333-369.

6. E.B. Bogomolny and J.P. Keating, Random matrix theory and the Riemann zeros I: three- and four-point correlations, Nonlinearity 8 (1995), 1115-1131.

7. P. Borwein, S. Choi, B. Rooney, and A. Weirathmueller, The Riemann Hypothesis: A Resource for the Afficionado and Virtuoso Alike, CMS Books in Mathematics (Springer, New York) (2008).

8. J.-F. Burnol, A lower bound in an approximation problem involving the zeros of the Riemann zeta function, Advances in Mathematics **170** (2002), 56–70.

9. S. Chowla, The Riemann Hypothesis and Hilbert's Tenth Problem, Mathematics and Its Applications, Vol. 4, Gordon and Breach Science Publishers, New York, (1965).

10. J. B. Conrey, More than two fifths of the zeros of the Riemann zeta function are on the critical line, J. Reine Angew. Math. **399** (1989), 1-26.

11. J. B. Conrey, The Riemann hypothesis, Notices American Mathematical Society **50** (2003), 341-353.

12. J.B. Conrey and N.C. Snaith, Correlations of eigenvalues and Riemann zeros, Communications in Number Theory and Physics, 2 (2008), 477-536.

13. J. Derbyshire, Prime Obsession (Joseph Henry Press, Washington, DC, 2003).

14. F.J. Dyson, Statistical theory of energy levels of complex systems, J. Math. Phys. 3 (1962) 140-156.

15. S.F. Edwards, P.W. Anderson, Theory of spin glasses. J. Phys. F Met. Phys. **5** (1975), 965–974.

16. L. Euler, Varia observationes circa series infinitas, Comm. Acad. Sci. Imp. Petropol. **9** (1744), 160-188

17. M. E. Fisher, The Nature of Critical Points (Lectures in Theoretical Physics vol VIIC) ed W Brittin (Boulder, CO: University of Colorado Press) (1965), 1-159.

18. P.J. Forrester, A.M. Odlyzko, Gaussian unitary ensemble eigenvalues and Riemann $\zeta$ function zeros: A nonlinear equation for a new statistic, Phys. Rev. E 54 (1996), R4493-4495.

19. P.J. Forrester, N. C. Snaith,J. J. M. Verbaarschot, Developments in random matrix theory, J. Phys. A: Math. Gen. **36** (2003) R1-R10.

20. X. Gourdon, The $10^{13}$ first zeros of the Riemann zeta function, and zeros computation at very large height, (2003),

http://numbers.computation.free.fr/Constants/Miscellaneouszetazeros1e13-1e24.pdf

21. R. B. Griffiths, First-order phase transition in spin-one Ising systems, Physica **33**, (1967) 689-690.

22. L. Guth and J. Maynard, New large value estimates for Dirichlet polynomials, (2024), arXiv2405.20552.

23. G. H. Hardy, The zeros of the function ζ(s) of Riemann, Comptes Rendus Hebdomadaires des Seances de l Academie des Sciences, **158** (1914), 1012-1014.

24. J.T. Ho, J. D. Litster, Magnetic equation of state of $CrBr_3$ near the critical point, Phys. Rev. Lett., **22** (1969), 603-606.

25. P. Hoever, W.F. Wolff, J. Zittartz, Random layered frustration models, Z. Phys. B - Condensed Matter **41** (1981) 43-53.

26. P. Hoever, J. Zittartz, Thermodynamics of some layered Ising models, Z. Phys. B - Condensed Matter **44** (1981) 129-134.

27. C.P. Hughes, J.P. Keating and N. O'Connell, Random matrix theory and the derivative of the Riemann zeta function, Proc. R. Soc. London A 456 (2000), 2611-2627.

28. A.E. Ingham. On the estimation of N(σ, T). Quart. J. Math. Oxford Ser., **11** (1940), 291–292.

29. E. Ising, Beitrag zur theorie des ferromagnetismus, Z Phys, **31** (1925), 253-258.

30. D. Kapitan, A. Korol, E. Vasiliev, P. Ovchinnikov, A. Rybin, E. Lobanova, K. Soldatov, Y. Shevchenko, V. Kapitan, Application of machine learning in solid state physics, Solid State Physics, **74** (2023), 1-65.

31. B. Kaufman, Crystal statistics II: Partition function evaluated by spinor analysis, Phys. Rev. **76** (1949), 1232-1243.

32. J. P. Keating, N. C. Snaith, Random matrix theory and ζ(1/2 + it), Commun. Math.

Phys. **214** (2000), 57-89.

33. A. Knauf, The number-theoretical spin chain and the Riemann zeroes, Commun. Math. Phys. **196** (1998), 703-731.

34. H.A. Kramers, G.H. Wannier, Statistics of the two-dimensional ferromagnet, Phys. Rev. **60** (1941), 252-262.

35. A. Languasco, A. Perelli, A. Zaccagnini, Explicit relations between pair correlation of zeros and primes in short intervals, J. Math. Analysis Appl., 394 (2012), 761-771.

36. T.D. Lee, C.N. Yang, Statistical theory of equations of state and phase transitions. IL Lattice gas and Ising model, Phys. Rev. **87** (1952), 410-419.

37. N. Levinson, More than one third of zeros of Riemann's zeta-function are on σ = l/2. Adv. Math. **13** (1974), 383-436.

38. B.C. Li and W. Wang, Influence of a new long-range interaction on the magnetic properties of a 2D Ising layered model by using Monte Carlo method, Chinese Journal of Physics, **87** (2024), 525-539.

39. B.C. Li and W. Wang, Exploration of dynamic phase transition of 3D Ising model with a new long-range interaction by using the Monte Carlo method, Chinese Journal of Physics, **90** (2024), 15-30.

40. A.R. Lima, M. A. de Menezes, Entropy-based analysis of the number partitioning problem, Phys. Rev. E **63** (2001), 020106(R).

41. F.X. Liu, L.-M. Duan, Computational characteristics of the random-field Ising model with long-range interaction, Phys. Rev. A **108** (2023), 012415.

42. Y. Liu, V.N. Ivanovski, C. Petrovic, Critical behavior of the van der Waals bonded ferromagnet $Fe_{3-x}GeTe_2$, Phys. Rev. B **96** (2017), 144429.

43. B.M. McCoy, In: Phase Transitions and Critical Phenomena. C. Domb, M.S. Green, (eds.) London, New York: Academic Press (1972).

44. M. L. Mehta, Random Matrices (Elsevier, Amsterdam, 2004).

45. S. Mertens, Phase transition in the number partitioning problem, Phys. Rev. Lett. 81 (1998), 4281-4284.

46. H. Montgomery, The pair correlation of zeros of the zeta function, Analytic Number Theory (Proc. Sympos. Pure Math., Vol. XXIV, St. Louis Univ., St. Louis, MO., 1972), American Mathematical Society Providence, (1973), 181-193.

47. M. Mézard, G. Parisi, M.A. Virasoro, Spin Glass Theory and Beyond (World Scientific, Singapore), (1987).

48. S. J. Miller and R. Takloo-Bighash, An Invitation to Modern Number Theory (Princeton University Press, Princeton, 2006).

49. G. Mussardo, The quantum mechanical potential for the prime numbers, (1997), arXiv:cond-mat/9712010.

50. S. Nagy, R. Paredes, J.M. Dudek, L. Dueñas-Osorio, M.Y. Vardi, Ising model partition-function computation as a weighted counting problem, Phys. Rev. E **109** (2024), 055301.

51. C. M. Newman, The GHS inequafity and the Riemann hypothesis, Constr. Approx. **7** (1991) 389-399.

52. L. Onsager, Crystal statistics I: A two-dimensional model with an order-disorder

transition, Phys. Rev. **65** (1944), 117-149.

53. G. Parisi, Infinite number of order parameters for spin-glasses, Phys. Rev. Lett. **43** (1979), 1754-1756.

54. B. Riemann, Uber die Anzahl der Primzahlen unter einer gegebenen Grösse, Monatsb. der Berliner Akad., (1859), 671–680.

55. H.C. Rosu, Quantum Hamiltonians and prime numbers, Mod. Phys. Lett. A 18 (2003), 1205–1213

56. A. Sander, L. Burgholzery, R. Wille, Towards Hamiltonian simulation with decision diagrams, IEEE Inter. Conf. Quantum Computing and Engineering, (2023), arXiv:2305.02337.

57. P. Schofield, J. D. Litster, J.T. Ho, Correlation between critical coefficients and critical exponents, Phys. Rev. Lett., **23** (1969), 1098-1102.

58. D. Schumayer and D.A.W. Hutchinson, Physics of the Riemann hypothesis, Rev. Mod. Phys. **83** (2011), 307-330.

59. D. Schumayer, B.P. van Zyl and D.A.W. Hutchinson, Quantum mechanical potentials related to the prime numbers and Riemann zeros, Phys. Rev. E **78 (**2008**)** 056215.

60. S.K. Sekatskii, On the Hamiltonian whose spectrum coincides with the set of primes, (2007), arXiv:0709.0364.

61. O. Shanker, Random matrices, generalized zeta functions and self-similarity of zero distributions, J. Phys. A: Math. Gen. **39** (2006), 13983-13997.

62. R.R.P, Singhal, M.E. Fisher, Short-range Ising spin glasses in general dimensions,

J. Appl. Phys. **63** (1988), 3994–3996.

63. M. Srednicki, Nonclassical degrees of freedom in the Riemann Hamiltonian, Phys. Rev. Lett. **107** (2011), 100201.

64. O. Suzuki, Z.D. Zhang, A method of Riemann-Hilbert problem for Zhang's conjecture 1 in a ferromagnetic 3D Ising model: trivialization of topological structure, Mathematics, **9** (2021), 776.

65. E. Wigner, Random matrices in physics, SIAM Rev. **9** (1967), 1-23.

66. M. Wolf, Will a physicist prove the Riemann hypothesis? Rep. Prog. Phys. **83** (2020), 036001.

67. W.F. Wolff, P. Hoever, J. Zittartz, Layered inhomogeneous Ising models with frustration on a square lattice, Z. Phys. B - Condensed Matter **42** (1981) 259-264.

68. F.Y. Wu, G. Rollet, H.Y. Huang, J. M. Maillard, C.-K. Hu, and C.-N. Chen, Directed compact lattice animals, restricted partitions of an integer, and the infinite-state Potts model, Phys. Rev. Lett. **76** (1996), 173-176.

69. H.J. Xu, S. Dasgupta, A. Pothen, A. Banerjee, Dynamic asset allocation with expected shortfall via quantum annealing, Entropy, **25** (2023), 541.

70. C.N. Yang, T.D. Lee, Statistical theory of equations of state and phase transitions. I. Theory of condensation, Phys. Rev. **87** (1952), 404-409.

71. Z.D. Zhang, Conjectures on the exact solution of three - dimensional (3D) simple orthorhombic Ising lattices, Phil. Mag. **87** (2007), 5309-5419.

72. Z.D. Zhang, N.H. March, Three-dimensional (3D) Ising universality in magnets and critical indices at fluid-fluid phase transition, Phase Transitions **84** (2011), 299-307.

73. Z.D. Zhang, Mathematical structure of the three-dimensional (3D) Ising model, Chinese Phys. B **22** (2013), 030513.

74. Z.D. Zhang, O. Suzuki, N.H. March, Clifford algebra approach of 3D Ising model, Advances in Applied Clifford Algebras, **29** (2019), 12.

75. Z.D. Zhang, Computational complexity of spin-glass three-dimensional (3D) Ising model, J. Mater. Sci. Tech. **44** (2020), 116-120.

76. Z.D. Zhang, O. Suzuki, A method of the Riemann-Hilbert problem for Zhang's conjecture 2 in a ferromagnetic 3D Ising model: topological phases, Mathematics, **9** (2021), 2936.

77. Z.D. Zhang, Topological quantum statistical mechanics and topological quantum field theories, Symmetry, **13** (2022), 323.

78. Z.D. Zhang, Mapping between spin-glass three-dimensional (3D) Ising model and Boolean satisfiability problems, Mathematics, **11** (2023), 237.

79. Z.D. Zhang, Lower bound of computational complexity of Knapsack problems, AIMS Math. **10** (2025), 11918-11938.

80. Z.D. Zhang, Exact solution of the three-dimensional (3D) $Z_2$ lattice gauge theory, Open Physics, 23 (2025) 20250215.

81. Z.D. Zhang, Two-dimensional Ising model with the next nearest interactions, Phys. Rev. E 113 (2026) 064111.

82. Z.D. Zhang, Relationship between spin-glass three-dimensional (3D) Ising model and traveling salesman problems, arXiv: 2507.01914.

83. R. Zivieri, Critical behavior of the classical spin-1 Ising model for magnetic systems, AIP Advances **12** (2022) 035326.